\documentclass{eas}
\usepackage{graphicx}
\def\kmsmpc{\ {\rm km~s^{-1} Mpc^{-1}}}

\def\hmsun{h^{-1} M_\odot}
\def \etal {{\it et al.\ }}

\def \ie {{\it i.e.\ } }
\def \eg{{\it e.g.\ }}
\def\apj{ApJ}
\def\apjl{ApJ Letters}
\def\mnras{MNRAS}

\newcommand {\rs} {$R_s$}
\newcommand {\rvir} {$R_{vir}$}

\begin{document}

\TitreGlobal{Mass Profiles and Shapes of Cosmological Structures}

\title{Constrained Simulations of Dark Matter Halos}
\author{Hoffman, Y.}\address{Racah Institute of Physics, Hebrew University,
Jerusalem 91904, Israel}
\author{Romano-D\'{\i}az, E.}\address{Racah Institute of Physics, Hebrew University,
Jerusalem 91904, Israel}
\author{Faltenbacher, A.}\address{Physics Department, University of California, Santa Cruz,  CA 95064, USA}
\author{Jones, D.}\address{Dep. of Physics \& Astronomy,
University of Kentucky, Lexington, KY 40506-0055, USA}
\author{Heller, C.}\address{Department of Physics,  Georgia Southern University,
  Statesboro, GA 30460, USA}
\author{Shlosman, I.}\address{Dep. of Physics \& Astronomy,
University of Kentucky, Lexington, KY 40506-0055, USA}

\runningtitle{Evolution of Halos }
\setcounter{page}{1}
\index{Author1, A.}
\index{Author2, B.}
\index{Author3, C.}

%
\begin{abstract}

The formation and structure of dark matter  halos is studied by  constrained   simulations. A series of experiments of the
formation of a $10^{12}\hmsun$ halo is designed to study the dependence
of the density profile on its merging history.
We find that the halo growth consist of several quiescent phases intermitted by violent events, with the density well approximated by the NFW
profile during the former phases. We find that (1) the NFW scale radius
\rs\  stays constant during the quiescent phase and grows abruptly during the
violent one. In contrast, the virial radius
grows linearly during the quiescent and abruptly during the violent
phases. (2) The central density stays unchanged during the quiescent
phase while dropping abruptly during the violent phase, and it does not reflect
the formation time of the halo.
(3) The clear separation of the evolution of an individual halo into  quiescent and violent phases implies that its entire evolution cannot be fitted by simple scaling relations.

 \end{abstract}

 \maketitle

%

\section{Introduction}

The problem of the formation and structure of dark matter (DM) halos
constitutes one of the outstanding challenges of theories of cosmic
structure formation.    N-body simulations of structure formation in a CDM cosmology find that the density profile of  virialized halos is well approximated by the so-called NFW profile (Navarro, Frenk \& White 1996).
Analytical efforts of understanding the collapse of DM halos and the emergence of the NFW profile have focused on one of the two extreme  scenarios. One is based on the spherical infall model (\eg\ Hoffman \& Shaham 1985) and the other on the merging scenario (\eg\ Syer \& White 1998). It has been shown that both
models can adequately reproduce the NFW profile.  Cosmological simulations show that halo formation in general proceeds via major mergers, yet   simulations   of halo formation  via monolithic collapse reproduce the  NFW structure as well. So, it is clear that a  basic theory of gravitational collapse that can unite the seeming diametrically opposed models needs to be formulated.  We have recently embarked on a series of constrained simulations aimed at shedding light on the origin and evolution of the NFW profile. The simulations are based on initial conditions constructed by means of constrained realizations of Gaussian fields 
(Hoffman \& Ribak 1991),  designed to run a series of controlled experiments of halo formation
(Romano-D\'{\i}az \etal\ 2005).

It is generally accepted that the 
evolution of DM halos proceeds in two phases, of rapid and slow
accretion (\eg\ Wechsler \etal\ 2002). It is also known that an NFW structure is quickly
established after the rapid phase and is preserved during the slow
accretion.  
The role played by major mergers ({\it i.e.} rapid accretion) in establishing the NFW profle  motivates us to conduct a set of numerical experiments in which the merging history is designed by constrained realizations, keeping their structure otherwise identical. Therefore any possible differences in the outcome of the simulations must be attributed to their merging histories.
Such controlled numerical experiments can be easily performed by   constrained simulations.  
We design a set of numerical
experiments in which a given halo of mass $~10^{12}\hmsun$ (where $h$
is the Hubble's constant in units of $100 \kmsmpc$) is constrained to
follow different merging histories.

\section{Models}

A set of five different models, \ie\ experiments, is designed here
to probe different merging histories of a
$10^{12}\hmsun$ halo in an $\Omega_0=0.3$ OCDM cosmology.
This halo is then
modified to have different substructure on different mass scales
and locations designed to collapse at different times. The spherical
top-hat model is used here to set the numerical value of the
constraints
and the collapse time of substructures. 
This is used only as a rough guide as the
various substructures are neither spherical nor isolated. Furthermore,
the few constraints used here do not fully control the
experiments. The nonlinear dynamics can in principle affect the
evolution in a way not fully anticipated from the initial
conditions. Even more important is the role of the random component of
the constrained realizations (Hoffman \& Ribak 1991).
The models are designed as follows: Model A (the benchmark model) is
based on two constraints. One is that of a $10^{12}\hmsun$ halo at the
origin.
This halo is
embedded in a region corresponding to a mass of
$10^{13}\hmsun$ in which the over-density is zero --- a region
corresponding to an unperturbed Friedmann model . These two constraints are imposed on all other models.
Model B
adds two substructures of mass $5\times 10^{11}\hmsun$ within the
$10^{12}\hmsun$ halo.
Model C further splits
each one of model B halos
into two $2.5\times 10^{11}\hmsun$ substructures.
Thus, the benchmark halo is modified to follow two major
mergers events on its way to virialization. Model D takes the Model A
and imposes six different small substructures of mass $10^{11}\hmsun$ scattered within the big halo.  Model E attempts to
simulate a more monolithic collapse
in which a nested set of constraints, located at the origin, is set on
a range of mass scales down to $M=10^{10}\hmsun$.  All models have been constructed with the same seed of the
random field. All the density constraints constitute $(2.5-
3.5)\sigma$ perturbations (where $\sigma^2$ is the variance of the
appropriately smoothed field), and were imposed on a cubic grid of $128^3$ grid of a co-moving length $L=4 Mpc/h$ .   
The main halo consists of $\approx 10^6$ particles and is simulated by   means of
the FTM code.

\section{Results}

The analysis of the simulations is based on the construction of a halos catalog and by fitting an NFW density profile to individual halos. The evolution of the NFW parameters of the primary halo in each model is presented here.  The evolution toward the final halo is studied by tracking  in
time the main branch that leads to this halo. The  picture that
emerges is that of a halo undergoing phases of  slow and ordered evolution
intermitted by episodes of rapid mass growth via collapse and
major mergers.
 These are referred to as the quiescent and violent  phases. The evolution of \rs, the virial radius $R_{vir}$ and the mean density within \rs\ ($\rho_s$) is presented in Fig. 1. The trends found here are all reproduced when the fitting is done by a  generalized NFW profile.

\begin{figure}[h]
   \centering
 \includegraphics[width=11cm]{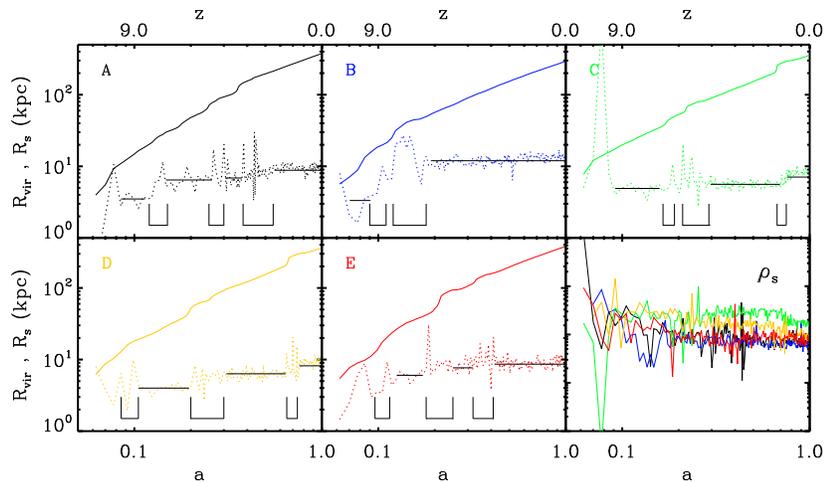}
      \caption{ Virial and scale radii behavior (continuous and dotted lines
respectively) as function of the expansion parameter $a$ for the five models.
The discontinuous
growths in \rs\ and \rvir\ match the violent phases that each halo passes through.
The horizontal bars represent the mean value of \rs\ within the quiescent phases.
The square brackets delineate the violent phases. The bottom right panel
shows the evolution of $\rho_{\rm s}$ (in arbitrary units) with colors corresponding to the models
in other panels.}
       \label{fig:fig1}
   \end{figure}

\section{Discussion}

The dense time sampling of the halos evolution enables us to  conclusively identify the trends in the evolution of the NFW parameters, some of which have been  just hinted  about in the literature so far. Here we focus on the evolution of \rs,  $R_{\rm vir}$ and the central density.
The main new results found here are:
(1) The NFW scale \rs\ stays constant during the quiescent
phases and changes abruptly during the violent ones. In contrast, $R_{\rm vir}$
is growing linearly in the quiescent and abruptly during the violent
phases;  
(2) The value of \rs\ reflects the violent merging history of the halo,
and it depends on the number of violent events and their fractional magnitudes,
independent of the time and order of these events; (3)  $\rho_{\rm s}$ stays unchanged during the quiescent
phases and drops abruptly during the violent phases. The corollary is
that $\rho_{\rm s}$ does not reflect the formation time of the halo;
(4) The relative change in \rs\ is a nonlinear function of the
relative absorbed kinetic energy within \rs\ in a violent event.
(5) The fact that the evolution of a given halo consists of a few quiescent phases intermitted by violent episodes implies that simple scaling relations can be applied only to a single
accretion trajectory but cannot be used to bridge
and extend over a few such trajectories.
We note, that the accretion trajectories in all models converge to the same
value. This is a reflection of the large-scale structure shared by
all the models and imposed by the constrained initial conditions.

The analogy between the halo evolution and thermodynamical processes has not
escaped our attention. Equating the quiescent phases with adiabatic
processes and the violent with non-adiabatic ones leads one
to associate the behavior of \rs\ with that of the entropy. In this
terminology the entropy remains constant in the quiescent phase and grows
discontinuously in the violent phase.
Also,  the accretion
trajectories play the role of  adiabats and the system jumps from one
adiabat to the other by a violent event, not unlike a shock wave.

The results obtained here pertain to the one halo studied
in the framework of an OCDM cosmology. Yet, the conclusions reached
from the set of experiments presented here are relevant to the
understanding of halo formation in the general CDM cosmologies and in
particular in the `benchmark' $\Lambda$CDM cosmology.


This research has been partially supported by the ISF -143/02  and the Sheinborn Foundation.


\end{document}